# Space-borne gravitational wave detectors as time-delayed differential dynamometers

Giuseppe Congedo, Rita Dolesi, Mauro Hueller, Stefano Vitale, and William J. Weber

*Department of Physics, University of Trento, and INFN*
*I-38123, Povo, Trento, Italy*

The basic constituent of many space-borne gravitational missions, in particular for interferometric gravitational waves detectors, is the so-called "link" made out of a satellite sending an electromagnetic beam to a second satellite. We illustrate how, by measuring the *time derivative of the frequency* of the received beam, the link behaves as a differential, time-delayed dynamometer in which the effect of gravity is exactly equivalent to an effective differential force applied to the two satellites. We also show that this differential force gives an integrated measurement of curvature along the beam. Finally, we discuss how this approach can be implemented to benefit the data analysis of gravitational wave detectors.

**PACS:** 04.80.Nn, 04.30.-w, 04.80.Cc,

With the establishment of laser interferometers as the state-of-the-art gravitational wave (GW) detectors, it has become customary to describe GWs as perturbations in the metric tensor that cause phase shifts of light beams exchanged by free test-particles. These phase shifts are indistinguishable from those caused, in flat space-time, by relative displacements of the test-particles, and are then treated as equivalent displacements. However GWs can equivalently be described as perturbations in the Riemann curvature tensor that cause tidal accelerations of free test-particles. This description, that was popular at the time of resonant detectors, has been abandoned, most likely because, with large detector baselines, light propagation delays make it less intuitive.

In this paper we show that the tidal force description is as simple as the displacement description and allows for treating each arm of an interferometric detector as an ordinary differential dynamometer performing time-delayed measurements. By differential dynamometer we mean here a device that measures the time-varying difference between the forces applied to two test-particles, by sources external to the dynamometer, along their joining line. This alternative viewpoint has practical consequences as it allows for a data processing approach that generates a time series consisting of combinations of (time-delayed) force, per unit mass, measurements [1], instead of relative displacements.

Force time-series have some useful properties. For instance, long-lived displacement transients related to the free dynamical evolution of the system disappear, as they are not caused by any external force. This is especially useful for space-borne detectors, where these transients may appear at the same time scale as gravitational wave signals. In addition, spurious force pulses appear as spikes within the force data series, while their effect on displacement data persists over the relaxation time of the detector. Static accelerations appear at dc, instead of causing quadratic runaway of data. Spurious forces that can be calculated from the independent measurement of some physical parameter (e.g. a magnetic force derived from the measurement of magnetic field gradient) can directly be subtracted from the data, without the need of filtering them with the transfer function of the detector. The same direct subtraction will be shown to apply also for known commanded forces and calculable elastic coupling effects that are relevant to systems employing both a satellite and an internal geodesic reference test mass, such as will be described shortly for space gravitational wave detection.

All in all these features produce a data series with a much-reduced dynamical range and with fewer systematics. The subtraction of these systematics from displacement data would require a substantial complication of the fitting templates and an increase in the required number of fitting parameters, including, for instance, the values of initial dynamical conditions. In the following we also discuss the practical guidelines to develop such an alternative data reduction approach, such as that currently being implemented for the LISA Pathfinder mission [2].

A more pronounced reduction of data dynamic range is obtained by data whitening, as is common, for instance, with ground-based GW detectors. However a generic whitening filter does not suppress the transients related to the free dynamical evolution of the system, unless it is explicitly designed as the series of a stage converting the data into force, followed by a stage that whitens the force data. An evaluation of the pros and cons of whitening force data, in order to achieve optimal dynamic range reduction, goes beyond the scope of this article. We note however that whitening would be difficult and even risky for the strongly signal dominated data expected from space-based detectors. The large gravitational wave signal power prevents the accurate noise calibration required by the whitening filter and does not allow for discriminating instrument noise against the expected GW foreground. Indeed proposed approaches to signal extraction with space-based GW detectors do not use data whitening and leave noise as one of the outputs of a global signal search procedure[3].

We now discuss the response of the basic constituent of a GW interferometric detector, and of other gravitational missions, a "link" composed of a satellite, that we call the "emitter" (e), sending a laser beam to a second satellite that we call the "receiver" (r). The (angular) frequency $\omega$ of the beam, with phase $\phi$ and wave vector $k_\mu$, is measured at both ends, and the Doppler shift, or difference of these two measurements

$$\omega_r - \omega_e \equiv c\left(\frac{d\phi}{d\tau_r} - \frac{d\phi}{d\tau_e}\right) = c\left(k_\mu^e v_e^\mu - k_\mu^r v_r^\mu\right) = ck_\mu^r\left(v_{e\to r}^\mu - v_r^\mu\right) \qquad 1$$

carries the physical information. Here "r" and "e" also indicate that measurements have been done at the event of reception and emission, respectively. We have used the notion that for any observer with velocity $v^\mu$,

$$\omega/c = d\phi/d\tau = \left(\partial\phi/\partial x^\mu\right)\left(dx^\mu/d\tau\right) = -k_\mu v^\mu \qquad 2$$

In the rightmost term of eq. 1 the symbol $e \to r$ indicates parallel transport along the beam from the emitter to the receiver, an operation that does not affect scalar products, while it converts $k_\mu^e$ into $k_\mu^r$. Finally notice that from 1 we get $\omega_r/\omega_e = d\tau_e/d\tau_r$.

In order to switch to acceleration and force, let's compare the derivatives of the frequency performed by the receiver and the emitter respectively, relative to their respective proper times $\tau_r$ and $\tau_e$

$$\frac{d\omega_r}{d\tau_r} - \frac{d\omega_e}{d\tau_e} = c\left[v_e^\mu \frac{Dk_\mu^e}{d\tau_e} + \frac{Dv_e^\mu}{d\tau_e}k_\mu^e - v_r^\mu \frac{Dk_\mu^r}{d\tau_r} - \frac{Dv_r^\mu}{d\tau_r}k_\mu^r\right] \equiv c\left[\left(G_e^\mu + F_e^\mu\right)k_\mu^e - \left(G_r^\mu + F_r^\mu\right)k_\mu^r\right] \qquad 3$$

Here we have used eq. 1 and the fact that ordinary and covariant derivatives are equal for a scalar. In addition we have used the geodesic equation to introduce the true four-forces (per unit mass) $F_r^\mu$ and $F_e^\mu$ acting on the receiver and the emitter respectively. Similarly we have defined an effective gravitational force (per unit mass) $G_e^\mu = \left(v_e^\mu/v_e^\rho k_\rho^e\right)\left(v_e^\nu Dk_\nu^e/d\tau_e\right)$ for the emitter and another, $G_r^\mu = \left(v_r^\mu/v_r^\rho k_\rho^r\right)\left(v_r^\nu Dk_\nu^r/d\tau_r\right)$, for the receiver.

In a GW detector true forces, which are part of noise, are small, and need only to be considered to first order and when not multiplied by factors <<1. Thus, in 3 one can approximate the wave vector with its flat space-time expression, and take $k_\mu^e \simeq k_\mu^r$ and $\omega_r \approx \omega_e$. In addition, as velocities are low, the scalar product between wave vector and four-forces reduces to the ordinary scalar product between their spatial parts. Then:

$$c\left(F_e^\mu k_\mu^e - F_r^\mu k_\mu^r\right) \approx \left(\omega_r/c\right)\hat{n}\cdot\left(\vec{F}_e - \vec{F}_r\right) \qquad 4$$

with $\vec{F}_r$ and $\vec{F}_e$ the ordinary forces (per unit mass) applied, respectively, to the emitter at the time of emission and to the receiver at the time of reception, and with $\hat{n}$ the unit three-vector parallel to the beam. This clarifies why such a system may be considered as a differential, time-delayed dynamometer: any differential force, with a component along the direction of the light beam, gives a signal $d\omega/dt$ at the receiver, in direct competition with possible gravitational or inertial effects.

Equation 3 does not allow for a direct calculation of the gravitational terms for a given physical situation. In order to clarify how these may be calculated, we expand more explicitly the scalar products $k_\mu^r G_r^\mu$ and $k_\mu^e G_e^\mu$. Let's first notice that $\tau_r$ effectively parameterizes the evolution of the beam geodesic. The coordinates along the beam received at time $\tau_r$ are $\xi(\lambda,\tau_r)$, where $\lambda$ is the affine parameter along the beam, and the wave vector $k_\mu(\lambda,\tau_r)$ is a function of both $\lambda$ and $\tau_r$ (FIG. 1).

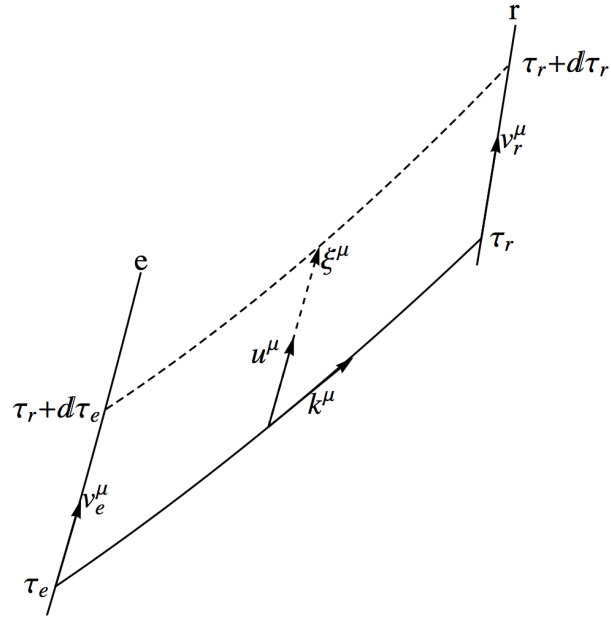

FIG. 1. Space-time sketch of the congruence.

This parameterization is similar but not identical to the standard null congruence discussed in textbooks [6]. Here $u^\mu(\chi,\tau_r) \equiv \partial\xi^\mu(\chi,\tau_r)/\partial\tau_r$ is not a null vector, and, in addition, due to the relative motion of receiver and emitter, the bounds of $\lambda$ may vary in time so that, even if we can always set $\lambda_e = 0$ at the emitter, at the receiver instead $\lambda_r = \lambda_r(\tau_r)$. In order to obtain a more practical congruence, we re-parameterize the beam by a new affine parameter $0 \leq \chi \leq 1$ and a "reduced" wave vector $\tilde{k}^\mu$ defined by $\chi = \lambda/\lambda_r(\tau_r)$ and $\tilde{k}^\mu \equiv \partial \xi^\mu/\partial \chi = \lambda_r k^\mu$ respectively. The most important properties of this congruence are:

$$\begin{aligned} u^\mu(\chi = 0, \tau_r) &= v_e^\mu(\tau_e)(d\tau_e/d\tau_r) \\ u^\mu(\chi = 1, \tau_r) &= v_r^\mu(\tau_r); \quad d\tilde{k}^\mu/d\tau_r = dv^\mu/d\chi \end{aligned} \qquad 5$$

By making explicit the covariant derivatives, we can then write that:

$$k_\mu^e G_e^\mu - k_\mu^r G_r^\mu = v_e^\mu \frac{Dk_\mu^e}{d\tau_e} - v_r^\mu \frac{Dk_\mu^r}{d\tau_r} \equiv v_e^\mu v_e^\nu \nabla_\nu k_\mu^e - v_r^\mu v_r^\nu \nabla_\nu k_\mu^r = v_e^\mu \frac{\omega_e}{\omega_r}(u^\nu \nabla_\nu k_\mu)_{\chi=0} - v_r^\mu (u^\nu \nabla_\nu k_\mu)_{\chi=1} \qquad 6$$

We can calculate this term by noticing that

$$(u^\nu \nabla_\nu k_\mu)_{\chi=1} - (u^\nu \nabla_\nu k_\mu)_{\chi=0} = \int_0^1 (du^\nu \nabla_\nu k_\mu/d\chi) d\chi = \int_0^1 \tilde{k}^\sigma \nabla_\sigma u^\nu \nabla_\nu k_\mu d\chi \qquad 7$$

It is straightforward to show, using the commutation properties of the congruence and the geodesic equation for the beam that

$$\tilde{k}^\sigma \nabla_\sigma u^\nu \nabla_\nu k_\mu = u^\nu \nabla_\nu \tilde{k}^\sigma \nabla_\sigma k_\mu + \tilde{k}^\sigma u^\nu R^\rho_{\sigma\nu\mu} k_\rho = \lambda_r k^\sigma u^\nu R^\rho_{\sigma\nu\mu} k_\rho \qquad 8$$

where $R^\rho_{\sigma\nu\mu}$ is the Riemann curvature tensor. Thus, finally:

$$\begin{aligned} k_\mu^e G_e^\mu - k_\mu^r G_r^\mu &= \\ &= \frac{Dk_\mu^r}{d\tau_r}\left(v_e^\mu \frac{\omega_e}{\omega_r} - v_r^\mu\right) - v_e^\mu \frac{\omega_e}{\omega_r} \int_0^{\lambda_r} k^\sigma u^\nu R^\rho_{\nu\sigma\mu} k_\rho d\lambda \end{aligned} \qquad 9$$

In order to clarify the meaning of the first term on the right hand side of 9, notice that in flat space-time $k_\mu^e = k_\mu^r$. In the non-relativistic limit we also get $\omega_r \approx \omega_e \equiv \omega$ and $d\tau_e \approx d\tau_r = dx^0$. Then in this limit:

$$\frac{Dk_\mu^r}{d\tau_r}\left[v_e^\mu(\omega_e/\omega_r) - v_r^\mu\right] \approx \frac{dk_\mu}{dx^0} \frac{d(\xi_e^\mu - \xi_r^\mu)}{dx^0} \approx \frac{dk_\mu}{dx^0} \frac{d[k^\mu(cL/\omega)]}{dx^0} = \frac{cL}{\omega} \frac{dk_\mu}{dx^0} \frac{dk^\mu}{dx^0} = \frac{\omega L}{c^2}|\hat{n}|^2 = \frac{\omega L}{c^2}\Omega^2 \qquad 10$$

To derive the result in 10 we have used the orthogonality between $k_\mu$ and its time derivative, and we have called L/c the light travel time over the link, and $\Omega$ the angular velocity at which the receiver sees $\hat{n}$ rotating. By comparing this term with 4, one can recognize the effect of the difference of centrifugal force between the emitter and the receiver in a frame that rotates to keep the line joining them fixed.

Notice that equations 3 and 9 do not contain approximations and are valid at any order in the metric deviation. They can then be used to calculate the response of the link *in any gravitational experiment using electromagnetic rays*, not just for GW detection with laser interferometers. Together they show that the link is at once a *differential dynamometer* and an instrument measuring the *space-time curvature integrated over the beam geodesic*. Thus, in a real situation within the solar system, it allows to calculate, for instance, both the effect of the secularly changing field due to solar system bodies, and the effect of GWs.

Let's calculate the curvature contribution for GW. As the quantity is a scalar, we can perform the calculation in any reference frame. We pick the transverse traceless frame [7], and we treat the case of a GW on a flat background where the sole non-zero components of the deviation $h_{\mu\nu}$ of the metric tensor from the flat one, are $h_{11} = -h_{22} = h_+\left[x^0 - x^3\right]$ and $h_{12} = h_{21} = h_\times\left[x^0 - x^3\right]$. To first order in the wave amplitudes $h_+$ and $h_\times$, the beam geodesic can be approximated with a space-time straight line, and four-velocities need only to be considered to $0^{th}$ order. This gives

$$-c\int_0^{\lambda_r} k^i R^j_{i00} k_j \, d\lambda = \omega \frac{1}{2(1-n_3)} \left\{ n_i n_j \left(\frac{dh_{ij}}{dx^0}\right)_{x_r^0 - x_r^3 - L(1-n_z)} - n_i n_j \left(\frac{dh_{ij}}{dx^0}\right)_{x_r^0 - x_r^3} \right\} \qquad 11$$

With $1 \leq i, j \leq 3$. The right hand side of eq.11 is the time derivative of the standard formula for the frequency shift along the outgoing path in a single interferometer link [8], with the time derivative calculated at the appropriate events. This shows that, within the linear approximation, "force" GW templates are just the first (delayed) derivatives of the corresponding frequency templates. In the same flat background, GW corrections to the centrifugal effect in 10 can be neglected as their leading term is of order $(\Omega/\omega_{GW})h \ll h$, where $\omega_{GW}$ is the angular frequency of the gravitational wave.

Let's now discuss how these results apply to a full space-borne GW detector [4], where, in its simplest configuration, a central satellite, nominally in free fall, sends two light beams to two other distant satellites also in free fall. Each distant satellite sends back a second beam, phase locked to the incoming one. Thus the detector consists of two pairs of phase-locked, counterpropagating links. In standard operations, the measured difference between the phases or the frequencies of the two returning links carries the GW signal. This differential scheme effectively suppresses the laser frequency noise [5], as this is common to both arms and does not require an absolute measurement of frequency with an accuracy that currently could not be achieved.

Eqs. 3 and 9 can be used sequentially, in a straightforward way, to generate the response of the two concatenated counter-propagating links in one arm. The output of the arm is the derivative of the frequency of the beam received at the central satellite. The outputs of the two arms, in the ideal Michelson configuration, can then directly be subtracted to generate the final signal, free of frequency noise. For the case of unequal arms the frequency derivatives can be time shifted and combined, to suppress laser frequency noise, into one of the combinations of the time delay interferometry [5]. It is straightforward to show that all these combinations maintain the basic results that GW signals, in the linear limit, are just the time-delayed time derivatives of the standard frequency signals.

Furthermore, in a real detector, the true force $\vec{F}_i$ on each satellite within a link, which would completely obscure GW signals, is estimated by measuring the satellite displacement relative to a free falling reference test-mass. This test mass is hosted inside the satellite and is well shielded against spurious forces. Test-mass to satellite displacements and satellite-to-satellite displacements are combined in an effective relative displacement of the two test-masses, which is nominally corrupted only by the effect of the much smaller spurious forces acting on the test-masses. We now illustrate how this correction is simplified and made immune to some systematics by working with forces instead of displacements.

The satellite and the test-mass form a two-body, controlled dynamical system. By controlled we mean here that for some degrees of freedom, including the relative motion of the test-mass and the satellite along the line of sight of the link, the system is stabilized by active control loops, often referred to as "drag-free control", pushing the satellite, via a set of thrusters, to follow the test-mass. Along the remaining degrees of freedom, control loops act instead, via electrostatic forces, on the test-mass to stabilize it relative to the satellite. The, linearized, non-relativistic dynamics, which is sufficient here, of this two-body system obeys

$$\ddot{Q}_i - M_{ik}^{-1} k_{kj} * q_j = F_i; \quad \ddot{q}_i + m_{ik}^{-1} k_{kj} * q_j = f_i^e + f_i^c - \ddot{Q}_i; \qquad 12$$

In equation 12 the $Q_i$ $(1 \le i \le 6)$ are the generalized coordinates, three displacements and three angles, of the satellite relative to the local inertial frame, and the $q_i$ are the generalized coordinates of the test-mass *relative to the satellite*, coordinates that in a real detector are measured, either by a dedicate interferometer or by capacitive sensors. The tensors $M_{ij}$ and $m_{ij}$ contain the mass and the moments of inertia of satellite and test-mass respectively, while the tensor $k_{ij}$ represents the effect of the static force gradient and of possible damping that are realistically assumed here to be generated by the satellite. The symbol $*$ indicates time convolution, as the action of these forces is not instantaneous. Furthermore $F_i$ is the force/torque (per unit mass/moment of inertia) acting on the satellite, including those applied by the drag-free control, so that the three first components coincide with the force vector discussed so far. Finally $f_i^c$ are the forces on the test-mass due to control loops while $f_i^e$ represent all the remaining forces acting on the test-mass. Notice that, due to imperfections and misalignments, the control forces $f_i^c$ have components along all degrees of freedom, including one along the direction of the link. Notice also that 12 has non-zero solutions also for $F_i = f_i^e = 0$, representing the free dynamical evolution of the system. From 12 it follows that:

$$\ddot{q}_i + \left(m_{ik}^{-1} + M_{ik}^{-1}\right) k_{kj} * q_j - f_i^c \equiv D_{ij} * q_j - f_i^c = f_i^e - F_i \qquad 13$$

Putting together 3, 4, 9, and 13, we conclude that, to first order in real forces, and if $D_{ij}$ has been calibrated with a proper set of dynamical measurements, one can generate a time series of effective difference of force per unit mass:

$$\Delta F_{\hat{n}} \equiv \frac{c}{\omega_r} \frac{d\omega_r}{dt_r} + n_i \left(D_{ij}^e * q_j^e - D_{ij}^r * q_j^r\right) + n_i \left(f_i^{c,r} - f_i^{c,e}\right) =$$
$$= \frac{c^3}{\omega_r} \left(G_e^\mu k_\mu^e - G_r^\mu k_\mu^r\right) + \frac{c}{\omega_r} \frac{d\omega_e}{dt_e} + \hat{n} \cdot \left(\vec{f}_e - \vec{f}_r\right) \qquad 14$$

with $t_r = \tau_r / c$ and $t_e = \tau_e / c$. In this way, we construct an estimator $\Delta F_{\hat{n}}$ — based on our measurements of the evolution of the time derivative of the measured receiver frequency $\omega_r$, the relative test-mass - satellite coordinates $q_i^e$ and $q_i^r$, and the applied forces $f_i^{c,e}$ and $f_i^{c,r}$ — that estimates the gravitational and inertial signal $\left(G_e^\mu k_\mu^e - G_r^\mu k_\mu^r\right)$. This estimate is corrupted by the final two noise terms created by the frequency noise of the emitter and the relative force (per unit mass) on the two reference test-masses, but is immune to the effect of dynamical transients, cross-talk of actuation forces etc.

The estimator $\Delta F_{\hat{n}}$ must also be calculated with signals $s_i$ from real displacement sensors, which contain both noise $s_i^n$ and imperfect mixing of different degrees of freedom, with $s_i = S_{ij} * q_j + s_i^n$, where $S_{ij}$ should be diagonal and memoryless in the absence of imperfections. Thus in reality the best approximation to $D_{ij}*q_j$ one can calculate is $D_{ij}*S_{jk}^{-1}*s_k$, thus requiring the calibration of both $D_{ij}$ and $S_{ij}$. In addition, data are corrupted not just by true forces but also by the equivalent force $D_{ij}S_{jk}^{-1}s_k^n$ due to the readout noise. This is, not surprisingly, equivalent to the displacement analysis, as no linear operation on the data can alter the signal to noise ratio. Finally, constructing the time series in eq. 14 requires the knowledge of the true control forces that have been applied to the test-masses. In reality the available information is only about the forces $\tilde{f}_i^{c,r}$ and $\vec{f}_i^{c,e}$ that have been *commanded* to the actuators. These are related to the former by $f_i^{c,r} = A_{ij}^r * \tilde{f}_j^{c,r}$, $f_i^{c,e} = A_{ij}^e * \tilde{f}_j^{c,e}$, where the operator $A_{ij}$ represents delays, cross talk, and inaccuracies of calibration. This operator needs also to be calibrated.

The calibration of all these operators is also needed when extracting GW signals from displacement time series. The difference here is that the operators are used to reduce the data *before* fitting GW templates to them, while in the classical approach they must be used to calibrate the templates.

It is also straightforward to use 14 to calculate the response of the complete interferometer, for which the result that the coupled free dynamics of the satellite and the test-mass is suppressed, is fully mantained.

The method of converting the data "at input", that is, of reducing them to force time series, is just an extension of a classical method adopted with other dynamical experiments like torsion pendulums [1]. It has also been adopted for LISA Pathfinder [9], a precursor of space-borne GW detectors developed by ESA, that carries a miniature version of an interferometric arm. In both the torsion pendulum experiments and in the simulation of LISA Pathfinder with the mission end-to-end simulator, we observe the accurate suppression of any large amplitude transient relaxation due to off-equilibrium initial conditions that instead dominate the displacement data. We also observe the resilience of the method to large force spikes that remain short-lived within the force data, while they appear as long relaxation transients within the displacement data series. Recent progress in this field with LISA Pathfinder will be reported in a separated paper.

As a final observation, we point out that the gravitational forces in eq. 14 also include the secular terms due to the quasi-static gravitational field in the solar system. Within force data, the amplitude ratio of these terms to the GW signals is suppressed, with respect to that in displacement data, by a factor of the order of the square of the ratio of their frequencies, that is, as a minimum, by a factor of order $(10^4/3\times10^7)^2 \approx 10^{-7}$, thus reducing the dynamical range of the data by the same factor. In addition these secular contributions can be calculated from the satellite tracking data and from the solar system ephemeris, and directly subtracted from the force data. The final accuracy with which this subtraction can be done depends on the effective performance of the tracking system. Accuracy reported for Doppler tracking of satellites [10] would likely allow to calculate and subtract these terms well below the residual very low frequency noise of the instrument due to thermal drift and other sources.

We thank Sergio Zerbini and Luciano Vanzo for many useful discussions. We also thank Philippe Jetzer for his reading of the manuscript. This work was supported by the Istituto Nazionale di Fisica Nucleare and the Agenzia Spaziale Italiana (LISA Pathfinder contract).